\documentstyle[12pt,epsf]{article}
\begin{document}

\title{ Relativistic motion with linear dissipation } \author{ G. Gonz\'alez  \thanks{E-mail:
gabriel${}_{-}$glez@yahoo.com},
\\{\it Departamento de Matem\'aticas y F{\'\i}sica} \\{\it
I.T.E.S.O.}\\ {\it Perif\'erico Sur \# 8585 C.P. 45090} \\
{\it Guadalajara, Jalisco, M\'exico.}}
\date{}
\maketitle

\begin{abstract}
A general formalism for obtaining the Lagrangian and Hamiltonian for a one dimensional dissipative system is developed. The formalism is illustrated by applying it to the case of a relativistic particle with linear dissipation. The relativistic wave equation is solved for a free particle with linear dissipation.\\ \\
{\em Key words:} Lagrangian, Hamiltonian, dissipative system.\\
{\em PACS:} 45.20.Jj, 03.65.Pm

\end{abstract}

\newpage

\section{Introduction}
It is very well known that for non dissipative systems the Lagrangian and Hamiltonian can be easily obtained by subtracting or adding respectively the kinetic and potential energy of the system \cite{B1}, but when we have dissipation in our dynamical system this construction is not useful and the corresponding Lagrangian and Hamiltonian is certainly not trivial to obtain \cite{GG,GL1}. The reason is that there is not yet a consistent Lagrangian and Hamiltonian formulation for dissipative systems. The problem of obtaining the Lagrangian
and Hamiltonian from the equations of motion of a mechanical
system is a particular case of ``The Inverse problem of the
Calculus of Variations" \cite{Sa,Vu}. This topic has been studied by
many mathematicians and theoretical physicists since the end of
the last century. The interest of physicists in this problem has
grown recently because of the quantization of dissipative
systems. A mechanical system can be quantized once its Hamitonian
is known and this Hamiltonian is usually obtained from a Lagrangian. 
The problem of quantizing dissipative systems has been extensively studied for nonrelativistic systems \cite{DHO,GG1} but little has been done for relativistic systems. \\
The main purpose of this paper is to develope a general formalism to obtain Lagrangians and Hamiltonians for one dimensional dissipative systems and apply it to the case of a relativistic particle under the action of a dissipative force which is proportional to its velocity. Once knowing the Hamiltonian of the system the relativistic wave function is obtained for a free particle with linear dissipation.

\section{Lagrangian and Hamiltonian for dissipative systems}
Newton's equation of motion for one dimensional systems
can be written as the following dynamical system
\begin{equation}
\frac{dx}{dt}=v, \qquad \frac{dv}{dt}=F(t,x,v), \label{eq1}
\end{equation}
where $t$ is time, $x$ is the position of the particle, $v$ is the velocity and
$F(t,x,v)$ is the force divided by the mass of the particle. If a Lagrangian function $L(t,x,v)$ is given for (\ref{eq1}) the Hamiltonian of the system can be obtained by the Legendre transform 
\begin{equation}
H(t,x,p)=pv(t,x,p)-L(t,x,v(t,x,p)),
\label{eq2}
\end{equation}
where $v(t,x,p)$ is the inverse function of the generalized linear momentum given by 
\[
p=\frac{\partial L}{\partial v}(t,x,v).
\] 
If the Lagrangian function has no explicit time dependence then the Hamiltonian of the system is a constant of motion \cite{B1}.\\
Assuming the following condition over the Hamiltonian
\begin{equation}
\frac{\partial^2 H}{\partial x \partial v}= \frac{\partial^2
H}{\partial v \partial x}, \label{eq3}
\end{equation}
then (\ref{eq3}) leads to
\begin{equation}
v\frac{\partial^3 L}{\partial x\partial v^2} + F\frac{\partial^3 L}{\partial
v^3}+\frac{\partial^3 L}{\partial v\partial t\partial v}+\frac{\partial F}{\partial v}\frac{\partial^2 L}{\partial v^2}=0, \label{eq4}
\end{equation}
where the Euler-Lagrange equation
\begin{equation}
\frac{d}{dt}\left(\frac{\partial L}{\partial v}\right)=\frac{\partial L}{\partial x},
\label{eq5}
\end{equation}
has been used. Therefore, in order to obtain the Lagrangian of (\ref{eq1}) we have to find a nontrivial solution for (\ref{eq4}) which for the general case is difficult to find \cite{Vu} but it can be less difficult if we consider cases where the generalized linear momentum $p=\partial{L}/\partial{v}$ has the following forms
\begin{description}
	\item [p=p(x,v)]: If the generalized linear momentum is independent of time, then equation (\ref{eq4}) turns into
	\begin{equation}
v\frac{\partial G}{\partial x} + F\frac{\partial G}{\partial
v}+\frac{\partial F}{\partial v}G=0
	\label{eq6}
	\end{equation}
where $G=\partial^2 L/\partial v^2$. The general solution for (\ref{eq6}) is given by
\begin{equation}
\frac{\partial^2 L}{\partial v^2}=\exp{\left(\int\frac{\partial F}{\partial v}\,dt\right)},
\label{eq7}
\end{equation}	
and the Lagrangian is obtained through the integration
\begin{equation}
L(x,v)=\int\,dv\int\,G(x,v)\,dv+f_{1}(x)v-f_{2}(x), \label{eq8}
\end{equation}
where $f_{1}(x)$ and $f_{2}(x)$ are arbitrary functions. The
second term on the right side of (\ref{eq8}) corresponds to a
gauge of the Lagrangian which brings about an equivalent
Lagrangian \cite{B1}, and it is possible to forget it.
	\item [p=p(t,v)]: If the generalized linear momentum is independent of position, then equation (\ref{eq4}) turns into
\begin{equation}
F\frac{\partial^3 L}{\partial v^3}+\frac{\partial^3 L}{\partial v\partial t\partial v}+\frac{\partial F}{\partial v}\frac{\partial^2 L}{\partial v^2}=0,
	\label{eq9}
	\end{equation}
which means that
\[
\frac{\partial}{\partial v}\left(F\frac{\partial^2 L}{\partial v^2}+\frac{\partial^2 L}{\partial t\partial v} \right)=0,
\]
the term in parenthesis is $dp/dt$, therefore if the generalized linear momentum is independent of position then the Euler-Lagrange equation must be of the form
\begin{equation}
\frac{dp}{dt}=f(t,x), \label{eq10}
\end{equation}
where $f(t,x)$ is the generalized force and may be any arbitrary function of position and time. Writing the generalized linear momentum as $p=\mu (t)g(v)$ then equation (\ref{eq10}) is given by
\begin{equation}
\frac{d\mu}{dt}g(v)+\mu(t)\frac{dg}{dv}\frac{dv}{dt}=f(t,x),
\label{eq11}
\end{equation}
if we are dealing with a relativistic system with linear dissipation, then
\begin{equation}
\frac{dv}{dt}=\left(-\frac{\partial U(t,x)}{\partial x}-\gamma v\right)(1-v^2/c^2)^{3/2},
\label{eq12}
\end{equation}
where $\gamma$ is a positive real parameter, $c$ is the speed of light and $U(t,x)$ represents the potential energy of the system. Substituting (\ref{eq12}) into (\ref{eq11}) we have
\begin{equation}
\frac{d\mu}{dt}g(v)+\mu(t)\frac{dg}{dv}\left(-\frac{\partial U(t,x)}{\partial x}-\gamma v\right)(1-v^2/c^2)^{3/2}=f(t,x),
\label{eq13}
\end{equation} 
if
\[
\frac{dg}{dv}=\frac{1}{(1-v^2/c^2)^{3/2}}
\]
then equation (\ref{eq13}) is given by
\begin{equation}
v\left(\frac{1}{\sqrt{1-v^2/c^2}}\frac{d\mu}{dt}-\gamma\mu(t)\right)-\mu (t)\frac{\partial U}{\partial x}=f(t,x),
\label{eq14}
\end{equation}
since equation (\ref{eq14}) must not depend on the velocity then 
\begin{equation}
\frac{1}{\sqrt{1-v^2/c^2}}\frac{d\mu}{dt}=\gamma\mu(t),
\label{eq15}
\end{equation}
the solution for (\ref{eq15}) is given by
\begin{equation}
\mu (t)=\exp{\left(\gamma\int\sqrt{1-v^2/c^2}\,dt\right)}=e^{\gamma\tau(t)},
\label{eq16}
\end{equation}
where
\begin{equation}
\tau(t)=\int\sqrt{1-v^2/c^2}\,dt,
\label{eq17}
\end{equation}
is known as the proper time of the particle \cite{B1}. Therefore equation (\ref{eq12}) can be expressed as
\begin{equation}
\frac{d}{dt}\left(\frac{ve^{\gamma\tau}}{\sqrt{1-v^2/c^2}}\right)=-e^{\gamma\tau}\frac{\partial U}{\partial x},
\label{eq18}
\end{equation}
and the Lagrangian and Hamiltonian for (\ref{eq18}) are given by
\begin{equation}
L(t,x,v)=-c^2 e^{\gamma\tau}\sqrt{1-v^2/c^2}-e^{\gamma\tau}U(t,x),
\label{eq19}
\end{equation}
\begin{equation}
H(t,x,p)=c^2 e^{\gamma\tau}\sqrt{1+p^2e^{-2\gamma\tau}/c^2}+e^{\gamma\tau}U(t,x).
\label{eq20}
\end{equation}
If $v<<c$ then (\ref{eq20}) reduces to the so called Caldirola-Kanai Hamiltonian
\begin{equation}
H(t,x,p)\approx \frac{p^2e^{-\gamma t}}{2}+e^{\gamma t}U(t,x)+c^2e^{\gamma t},
\label{eq21}
\end{equation}
which describes the motion of a nonrelativistic particle with linear dissipation \cite{DHO}.
\end{description}

\section{Relativistic wave equation with dissipation}
In all the subsequent analysis we will set $c=\hbar=1$.
Let us now consider the wave equation for a free relativistic particle subject to a dissipative force which is proportional to its velocity, the equation of motion of this system is given by 
\begin{equation}
\frac{dv}{dt}=-\gamma v(1-v^2)^{3/2}
\label{eq22}
\end{equation}
and the Hamiltonian for (\ref{eq22}) is
\begin{equation}
H(t,x,p)= e^{\gamma\tau}\sqrt{1+p^2e^{-2\gamma\tau}}
\label{eq23}
\end{equation}
making the usual substitution $p\rightarrow -i\partial/\partial x$ and $H\rightarrow i\partial/\partial t$ and letting equation (\ref{eq23}) act on a wave function $\Psi(t,x)$ we obtain the following relativistic wave equation
\begin{equation}
\left(\frac{\partial^2}{\partial t^2}-\frac{\partial^2}{\partial x^2} +e^{2\gamma\tau}\right)\Psi=0,
\label{eq24}
\end{equation}
Looking for solutions of the form $\Psi(t,x)=f(t)\psi(x)$ we obtain the following equations for $f(t)$ and $\psi(x)$
\begin{equation}
\frac{d^2 f}{dt^2}+\left(k^2+e^{2\gamma\tau}\right)f=0, \qquad \frac{d^2 \psi}{dx^2}+k^2\psi=0,
\label{eq25}
\end{equation}
where $k^2$ is the separation constant. Therefore 
\begin{equation}
\psi(x)=e^{\pm ikx}.
\label{eq26}
\end{equation}
So far, everything is exact, but to obtain the solution for $f(t)$ we have first to specify $\tau (t)=\int\sqrt{1-v^2}\,dt$, which can be done integrating the equation of motion (\ref{eq22}), doing this we get
\begin{equation}
-\gamma t=\frac{1-\xi\tanh^{-1}\xi}{\xi},
\label{eq27}
\end{equation}
where $\xi=\sqrt{1-v^2}$.
Expanding the term $\tanh^{-1}\xi$ and taking into account only terms less than or equal to $\xi^3$, then equation (\ref{eq27}) turns into
\begin{equation}
\xi^2-\gamma t \xi-1=0,
\label{eq28}
\end{equation}
solving equation (\ref{eq28}) we have
\begin{equation}
\xi(t)=\frac{\gamma t}{2}\left(1\pm\sqrt{1+\left(\frac{2}{\gamma t}\right)^2}\right),
\label{eq30}
\end{equation}
we have two solutions for $\xi(t)$, but we can get rid of one knowing that proper time runs slowly than coordinate time \cite{B1}, therefore we must choose
\begin{equation}
\xi(t)=\frac{\gamma t}{2}\left(1-\sqrt{1+\left(\frac{2}{\gamma t}\right)^2}\right).
\label{eq31}
\end{equation}
If $2<\gamma t$, then 
\begin{equation}
\xi(t)=\frac{\gamma t}{2}\left(1-1-\frac{1}{2}\left(\frac{2}{\gamma t}\right)^2-\frac{1}{8}\left(\frac{2}{\gamma t}\right)^4-\cdots\right),
\label{eq32}
\end{equation}
therefore
\begin{equation}
\tau(t)=\int\xi(t)\,dt\approx -\frac{\ln(t)}{\gamma},
\label{eq33}
\end{equation}
equation (\ref{eq33}) is valid only for $2/\gamma<t<1$. Substituting (\ref{eq33}) into (\ref{eq25}) we have
\begin{equation}
\frac{d^2 f}{dt^2}+\left(k^2+\frac{1}{t^2}\right)f=0,
\label{eq34}
\end{equation}
equation (\ref{eq34}) represents a Bessel equation in its normal form with the following solution 
\begin{equation}
f(t)=\sqrt{t}\left(J_{i\sqrt{3}/2}(k t)+Y_{i\sqrt{3}/2}(k t)\right),
\label{eq35}
\end{equation}
where $J$ and $Y$ are Bessel's function of the first and second kind respectively \cite{W1}. Therefore the approximate wave function for a free relativistic particle under a force which is proportional to its velocity is given by
\begin{equation}
\Psi(t,x)=\sqrt{t}\, e^{ikx}\left(J_{i\sqrt{3}/2}(k t)+Y_{i\sqrt{3}/2}(k t)\right),
\label{eq36}
\end{equation} 
where $2/\gamma<t<1$.

\section{Conclusions}
A general formalism to obtain the Lagrangian and Hamiltonian for one dimensional dissipative systems was obtained. The Lagrangian and Hamiltonian for a relativistic particle with linear dissipation was deduced, using this Hamiltonian the relativistic wave function was obtain for a free particle with linear dissipation.


\newpage

\begin{thebibliography}{99}

\bibitem {DHO} Chung-In Um, Kyu-Hwang Yeon, Thomas F. George, (2002). {\it The quantum damped harmonic oscillator},
Physics Reports {\bf 362},  63-192.

\bibitem {B1} Goldstein H., (1980). {\it Classical Mechanics},
Addison-Wesley.

\bibitem{GG}  Gonz\'alez, G.  (2004). {\it Lagrangians and Hamiltonians for one dimensional autonomous systems},  Int. Jou. Theo. Phys. {\bf 43}, 1885-1890.

\bibitem{GL1} L\'opez, G.  (1996). {\it One dimensional autonomous systems and dissipative systems}, Ann. Phys. {\bf 251}, 363-383.

\bibitem{GG1}  L\'opez, G., Gonz\'alez, G.  (2004). {\it Quantum bouncer with dissipation},  Int. Jou. Theo. Phys. {\bf 43}, 1999-2008.

\bibitem {Sa} Santilli, R. M. (1978). {\it Foundations of Theoretical Physics I}, Springer-Verlag.

\bibitem {Vu} Vujanovic, B. D., Jones, S. E. (1989).
 {\it Variational methods in nonconservative phenomena}, Academic Press, Inc.

\bibitem {W1} Watson G. N., (1966). {\it A Treatise on the Theory of Bessel Functions},
Cambridge University Press.

\end{thebibliography}
\end{document}